**Title:** High-dimensional optical quantum logic in large operational spaces

**Authors:** Poolad Imany,[1,2,*] Jose A. Jaramillo-Villegas,[1,2,3] Mohammed S. Alshaykh,[1,2] Joseph M. Lukens,[4] Ogaga D. Odele,[1,2] Alexandria J. Moore, [1,2] Daniel E. Leaird,[1,2] Minghao Qi,[1,5] Andrew M. Weiner[1,2,5]

[1]School of Electrical and Computer Engineering, Purdue University, West Lafayette, IN, USA.

[2]Purdue Quantum Science and Engineering Institute, Purdue University, West Lafayette, IN, USA.

[3]Facultad de Ingenierías, Universidad Tecnológica de Pereira, Pereira, RIS, Colombia.

[4]Quantum Information Science Group, Oak Ridge National Laboratory, Oak Ridge, TN, USA.

[5]Birck Nanotechnology Center, Purdue University, West Lafayette, IN, USA.

[*]Correspondence: pimany@purdue.edu

**Abstract.** The probabilistic nature of single-photon sources and photon-photon interactions encourages encoding as much quantum information as possible in every photon for the purpose of photonic quantum information processing. Here, by encoding high-dimensional units of information (qudits) in time and frequency degrees of freedom using on-chip sources, we report deterministic two-qudit gates in a single photon with fidelities exceeding 0.90 in the computational basis. Constructing a two-qudit modulo SUM gate, we generate and measure a single-photon state with non-separability between time and frequency qudits. We then employ this SUM operation on two frequency-bin entangled photons—each carrying two 32-dimensional qudits—to realize a four-party high-dimensional Greenberger-Horne-Zeilinger state, occupying a Hilbert space equivalent to that of 20 qubits. Although high-dimensional coding alone is ultimately not scalable for universal quantum computing, our design shows the potential of deterministic optical quantum operations in large encoding spaces for practical and compact quantum information processing protocols.

**Introduction**

Quantum information processing has drawn massive attention due to its power in solving some crucial algorithms exponentially faster than their classical counterparts [1], as well as its ability to transmit information in a fully secure fashion, two capabilities looked to be combined in the emerging quantum internet [2]. Amongst the platforms that can exhibit quantum behavior, optical states have the advantages of low decoherence and suitability for long-distance communications, yet the weak coupling of photons to their surroundings also makes it extremely difficult to manipulate the state of one photon based on the state of another. This operation, needed for a two-qubit gate, is probabilistic with standard linear optics and photon counting [3]. Quantum gates have been demonstrated in a number of different photonic degrees of freedom such as polarization [4], orbital angular momentum [5], time [6], and frequency [7,8], and to sidestep the challenges of probabilistic multiphoton interactions, encoding qubits in different degrees of freedom (DoFs) in a single photon has been demonstrated, where each DoF carries one qubit and, now, operations between different qubits can be made deterministic [9,10]. This scheme allows encoding more quantum information in single photons, and can find use in stand-alone processing tasks or be subsequently incorporated into larger systems built on true photon-photon interactions, thus offering a potentially more efficient method for photonic quantum information processing. Even though in this case two and three-qubit operations can be executed with unity success probability, each DoF contains only one qubit, and the number of a photon's DoFs are limited; thus the size of the Hilbert space in which these deterministic transformations can happen is fairly moderate (e.g., an eight-dimensional Hilbert space has been demonstrated by encoding three qubits in three different DoFs of a single photon [10]).

In this article, we take advantage of the high dimensionality in two particular DoFs of a single photon—namely, time and frequency, which are both compatible with fiber optical transmission—to encode one *qudit* in each DoF. We consider multiple time bins and frequency bins; as long as the frequency spacing between different modes ($\Delta f$) and the time-bin spacing ($\Delta t$) are chosen such that they exceed the Fourier transform limit (i.e., $\Delta f \Delta t > 1$), we are able to manipulate the time and frequency DoFs independently in a hyper-encoding fashion, using concepts developed in time-division and wavelength-division multiplexing, respectively [11,12]. In other words, each time-frequency mode pair constitutes a well-defined entity, or plaquette [11,12], which is sufficiently separated from its neighbors to provide stable encoding (Fig. 1). Alternatively, this can be understood by considering bandwidth-limited plaquettes with individual spectral linewidth $\delta f$ (corresponding to temporal duration $\sim 1/\delta f$). These will not overlap in time-frequency space as long as the chosen bin separations satisfy $\Delta f > \delta f$ and $\Delta t > 1/\delta f$. Combined, then, these two equations yield the aforementioned condition $\Delta f \Delta t > 1$. An analogous process is at work in the advanced optical modulation formats gaining adoption in modern digital communications, where many bits are encoded in a single symbol via modulation of canonically conjugate quadratures [13]. Since our single photons can potentially be generated in a superposition of many time and frequency bins, multiple qubits can be encoded in each DoF, making our proposed scheme a favorable platform for deterministic optical quantum information processing on Hilbert spaces dramatically larger than previously demonstrated deterministic qubit-based gates. Ultimately, the total number of DoFs carried by a single photon is limited, so one cannot increase the Hilbert space indefinitely by encoding in increasingly more properties within individual photons. The Hilbert space can be increased, though, by expanding the dimensionality within each DoF. While enabling only linear scaling of the Hilbert space with the number of modes [14], and thereby not facilitating the exponential scaling required for fault-tolerant quantum computing, qudit encoding promises significant potential in the current generation of quantum circuits. It has been shown, for example, that two-qudit optical gates are useful in transmitting quantum states with higher information content per photon by means of qudit teleportation [15], a task that requires two-qudit gates which can operate on the different degrees of freedom of a single photon [16,17]—precisely the functionality we demonstrate here.

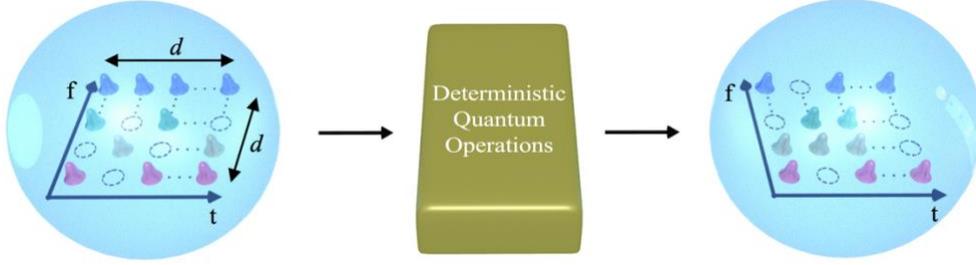

**Fig. 1.** Illustration of the scheme. Two qudits encoded in d time bins and frequency bins in a single photon, going through a deterministic quantum process. The single-photon can be encoded in an arbitrary superposition of different time and frequency bins; the unused time-frequency slots are shown with dashed circles. After the deterministic quantum process operates on the two-qudit state, the orientation of the time-frequency superpositions change to a new two-qudit state.

**Results**

To enable the realization of all single-qudit unitaries, it is sufficient to demonstrate the generalized Pauli gates X (cyclic shift) and Z (state-dependent phase), which are universal for single-qudit operations [5], and from which all $d$-dimensional Weyl operators can be constructed [18]. The Z gate applies a unique phase shift to each of the $d$ basis states, which can be easily executed with a phase modulator and a pulse shaper in the time domain and frequency domain, respectively. Specifically, for the basis state $|n\rangle$ ($n = 0,1,...,d-1$), we have $Z|n\rangle = \exp(2\pi i n/d)|n\rangle$ and $X|n\rangle = |n \oplus 1\rangle$, where $\oplus$ denotes addition modulo $d$. An X gate in the frequency domain can be realized using a Z gate sandwiched between two high-dimensional DFT gates. Such a DFT operation has been recently demonstrated [7], completing in principle the universal gate set for single-qudit frequency-domain operations. To complete the gate set in the time domain, we demonstrate the time-bin X gate presented in Fig. 2a, operating on time bins in three dimensions, a process which corresponds to state-dependent delay. Because the gate operates on each photon individually, we can simulate its performance with coherent states; the statistics of the input field have no impact on the principle of operation. Of course, to apply this gate in multiphoton quantum information processing, true single photons would need to be tested as well, the preparation or heralding of which is technically demanding and could introduce additional noise. However, as this noise is extrinsic to the gate itself, we focus on weak coherent states for initial characterization here. To test for the correct modal operation of this gate, we use a continuous-wave (CW) laser and prepare the desired weak coherent state by carving out three time bins $\{|0\rangle_t, |1\rangle_t, |2\rangle_t\}$ using an intensity modulator and manipulating their relative phases with a phase modulator. The time bins are 3 ns wide with $\Delta t = 6$ ns center-to-center spacing. To perform the X operation, we need to separate the time bins $|0\rangle_t$ and $|1\rangle_t$ from $|2\rangle_t$ and delay the route for time bins $|0\rangle_t$ and $|1\rangle_t$ by 3 bins (18 ns). We realize the necessary spatial separation between the time bins with a Mach-Zehnder modulator (MZM) switch. We emphasize that while most MZM designs are one-port devices, with one of the two output paths terminated, this 1×2 version permits access to both interferometer outputs, and accordingly it is in principle lossless—as required for a unitary operation. (In practice, of course, insertion loss reduces throughput, but it should be possible in the future to significantly reduce this loss through, e.g., on-chip integration.) After the path-dependent delay, another 1×2 MZM, but operated in reverse, can be used to recombine the time bins deterministically as well. However, due to lack of equipment availability, in this proof-of-principle experiment we employ a 2×2 fiber coupler for recombination, which introduces an additional 3 dB power penalty. To measure the gate output, we synchronize a single-photon detector and time interval analyzer with the generated time bins. The transformation matrix performed by the X gate when probed by single time bins yields a computational basis fidelity $\mathcal{F}_C$ of $0.996 \pm 0.001$, shown in Fig. 2b (see Methods). As such computational-basis-only measurements do not reflect the phase coherence of the operation, we next prepare superposition states as input and interfere the transformed time bins after the gate with a cascade of 1-bin and 2-bin delay unbalanced interferometers. In order to combat environmentally induced phase fluctuations, we stabilize both these interferometers and the X gate by sending a CW laser in the backwards direction and using a feedback phase control loop. We apply a phase of 0, $\phi$ and $2\phi$ to the time-bins $|0\rangle_t$, $|1\rangle_t$ and $|2\rangle_t$, respectively, with the phase modulator in the state preparation stage and sweep $\phi$ from 0 to $2\pi$, obtaining the interference pattern shown in Fig. 2c. After subtraction of the background, we calculate a visibility of $0.94 \pm 0.01$ from the maximum and minimum points, showing strong phase coherence (the ability to preserve and utilize coherent superpositions) between the time bins after the gate. If for concreteness we assume a channel model consisting of pure depolarizing (white) noise [18], we can use this visibility to estimate the process fidelity $\mathcal{F}_P$, finding $\mathcal{F}_P = 0.92 \pm 0.01$ for the X gate (see Methods). Given the ability to perform arbitrary one-qudit operations using combinations of X and Z gates, it follows that it is in principle possible to generate and measure photons in all mutually unbiased bases [19]—an essential capability for high-dimensional quantum key distribution (QKD) [20], which has been proven to offer greater robustness to noise compared to qubit-based QKD [21] and can enable significantly higher secret key rates over metropolitan-scale distances [22].

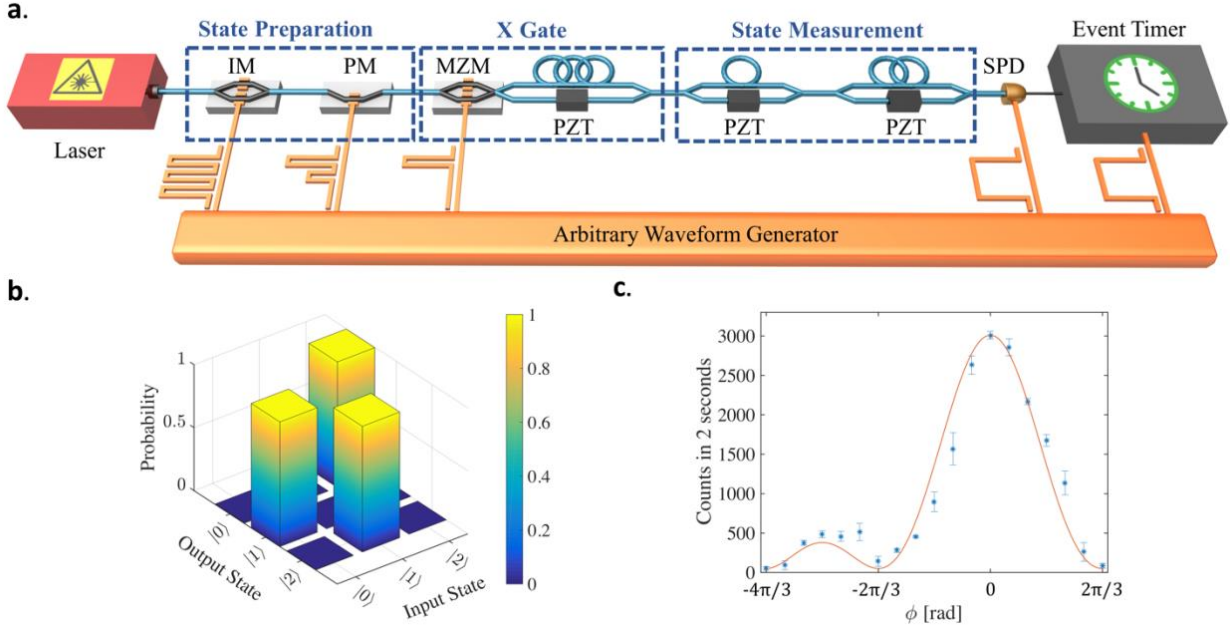

**Fig. 2. a**, Experimental setup of the state preparation, the X gate, and the state measurement. IM: intensity modulator. PM: phase modulator. MZM: Mach-Zehnder modulator PZT: piezo-electric phase shifter. SPD: single-photon detector. The circle-shaped fibers indicate the delay; each circle is equivalent to one time-bin delay (6 ns). **b**, The transformation matrix. **c**, Counts measured after overlapping all three output time bins, for a time-bin superposition state input into the X gate. The blue errorbars are obtained from 5 measurements for each phase. The subtracted background was about 200 per 2 seconds.

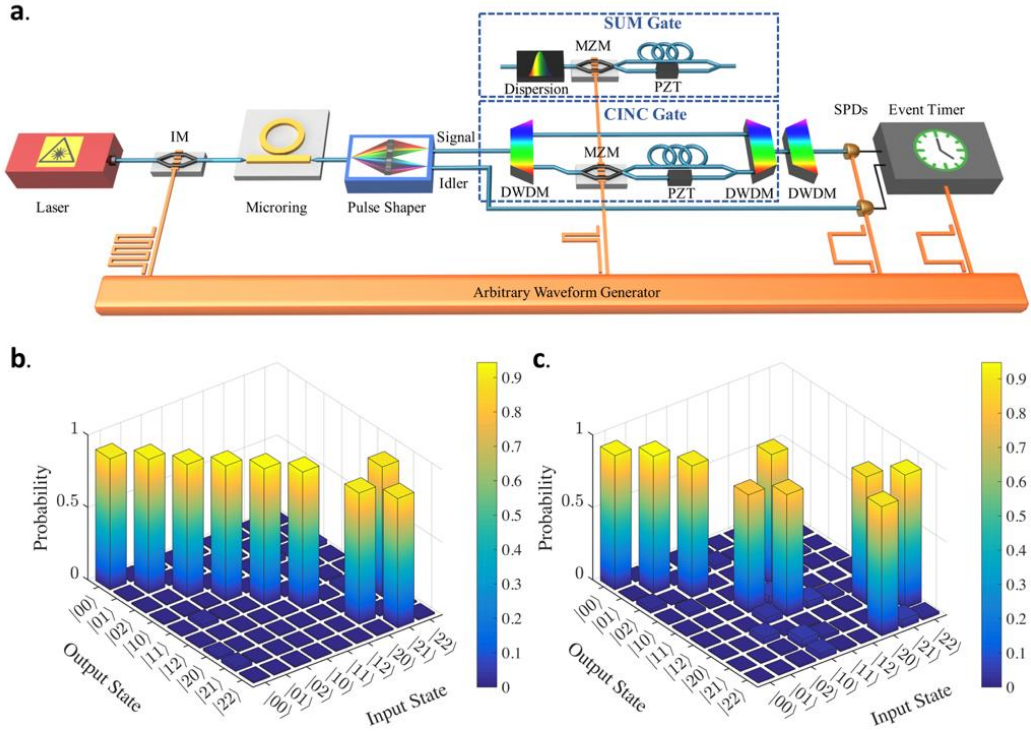

**Fig. 3. a**, Experimental setup for the CINC and SUM gate. The MZM for the CINC gate is driven such that it separates the time bin $|2\rangle_t$ from time bins $|0\rangle_t$ and $|1\rangle_t$. For the SUM gate, the MZM separates the time bins that fall outside of the computational space ($|3\rangle_t$ and $|4\rangle_t$) from the computational space time bins ($|0\rangle_t$, $|1\rangle_t$ and $|2\rangle_t$). DWDM: dense wavelength-division multiplexer. **b** and **c**, The experimental transformation matrix of the CINC and SUM gate, respectively. The accidentals were subtracted in the transformation matrices, and the coincidence to accidentals ratio was ~3.7 in the CINC and ~3 in the SUM case.

With this high-performance time-bin X gate in hand, we are then in a position to incorporate it into a frequency network to realize deterministic two-qudit gates, where the frequency DoF acts as the control and the time DoF is the target qudit. For this demonstration, instead of a weak coherent state, we utilize true single photons, heralded by detecting the partner photon of a frequency-bin entangled pair generated through spontaneous four-wave mixing in an on-chip silicon nitride microresonator. The time bins, defined by intensity modulation of the pump, couple into a microring resonator with a free spectral range (FSR) $\Delta f = 380$ GHz and resonance linewidths $\delta f \simeq 250$ MHz, generating a biphoton frequency comb. The time-bin and frequency-bin entanglement of such sources have been proven recently [23–26]. As our time- and frequency-bins exceed the Fourier limit ($\Delta f \Delta t = 2280$, $\delta f \Delta t = 1.5$), our time-frequency entangled photons can be considered hyper-entangled—that is, entangled in two fully separable DoFs. The signal and idler photons from the first three comb line pairs are then selected and separated with a commercial pulse shaper, as shown in Fig. 3a. Now that the time bins and frequency bins are all generated in the state preparation stage, the idler photons are sent to a single-photon detector for heralding, and the signal photons are what carry the two qudits in the three time bins $\{|0\rangle_t, |1\rangle_t, |2\rangle_t\}$ and frequency bins $\{|0\rangle_f, |1\rangle_f, |2\rangle_f\}$. This procedure lets us prepare any time-bin/frequency-bin product state $|m\rangle_t|n\rangle_f$ ($m, n = 0,1,2$) of the full computational basis set. In principle, we could also herald arbitrary time-frequency superposition states in this setup, by first sending the idler photon through a combination of time- or frequency-bin interferometers prior to detection in the temporal and spectral eigenbases. This more general case would permit the preparation of any two-qudit state and is an important area for further research.

As the first two-qudit gate, we demonstrate the controlled-increment (CINC) operation, where an X gate is applied to the time-bin qudit only when the frequency qudit is in the state $|2\rangle_f$. This two-qudit gate along with arbitrary single-qudit gates [which, as noted above, can be formed from qudit X and Z operations [5]] complete a universal set for any quantum operation [27]. To implement this gate, we separate $|2\rangle_f$ from the other two frequency bins with a dense wavelength-division multiplexing (DWDM) filter and route it to a time-bin X gate (Fig. 3a); no operation happens on the route of the other two frequency bins. The frequency bins are then brought back together with another DWDM with zero relative delay to complete the two-qudit gate operation. To measure the transformation matrix of this gate in the computational basis, we prepare the input state in each of the 9 combinations of single time bins and frequency bins, using the first intensity modulator and the pulse shaper, respectively. We then record the signal counts in all possible output time-bin/frequency-bin pairs, conditioned on detection of a particular idler time-frequency mode, by inserting three different DWDMs in the path of the signal photons to pick different frequency bins. The measured transformation matrix is shown in Fig. 3b, with accidental-subtracted fidelity $\mathcal{F}_C = 0.90 \pm 0.01$ (see Methods).

For the next step, we implement an even more complex operation, the SUM gate—a generalized controlled-NOT gate [28]—which adds the value of the control qudit to the value of the target qudit, modulo 3. In this gate, the time bins associated with $|0\rangle_f$ are not delayed, the time bins associated with $|1\rangle_f$ experience a cyclic shift by 1 slot, and the time bins corresponding to $|2\rangle_f$ go through a cyclic shift of 2 slots. To delay the time bins dependent on their frequencies, we induce a dispersion of -2 ns.nm$^{-1}$ on the photons using a chirped fiber Bragg grating (CFBG); this imparts 6 ns (1-bin) and 12 ns (2-bin) delays for the temporal modes of $|1\rangle_f$ and $|2\rangle_f$, respectively, as required for the SUM operation. However, this delay is linear—not cyclic—so that some of the time bins are pushed outside of the computational space, to modes $|3\rangle_t$ and $|4\rangle_t$. Returning these bins to overlap with the necessary $|0\rangle_t$ and $|1\rangle_t$ slots can be achieved using principles identical to the time-bin X gate with a relative delay of three bins. The experimental setup is shown in Fig. 3a, where we use the same techniques as for the CINC gate to measure the transfer matrix shown in Fig. 3c, with $\mathcal{F}_C = 0.92 \pm 0.01$. The fact that this SUM gate is implemented with qudits in a single step potentially reduces the complexity and depth of quantum circuits in algorithms that require an addition operation [29]. Lack of frequency shifting components in these gates can be confirmed by off-diagonal terms in Figs. 3b,c for which the input and output frequency bins differ.

To show the ability of our design to operate on large Hilbert spaces, we extend the dimensions of our qudits and encode two 16-dimensional quantum states in the time and frequency DoFs of a single photon. For this demonstration, as we want to use more time bins and set a smaller frequency spacing between modes, we use a periodically poled lithium niobate (PPLN) crystal as a broadband source of time-frequency entangled photons followed by a programmable pulse shaper to set the frequency spacing and linewidth, instead of a microring with fixed frequency spacing. (We note that, in principle, one could still use an integrated source for these experiments by appropriately engineering a microring's FSR, bandwidth, and resonance linewidth to realize spectral and temporal spacings tighter than the integrated photon sources currently available to us.) In this experiment, we first shine a 773 nm CW laser on the PPLN crystal, generating entangled photons with a bandwidth of ~ 5 THz [30]. We then carve 16 time bins with a full width at half maximum of ~200 ps and 1.2 ns spacing between them, to generate the time-bin qudits. Then, a pulse shaper is used to carve out the frequency of these entangled photons to generate sixteen 22 GHz wide frequency bins on both the signal and idler side of the spectrum, each spaced by 75 GHz from each other. Now that we have 16-dimensional qudits in both time and frequency, we send a heralded signal photon into the same SUM-gate structure. We note that after the CFBG, the individual time bins will spread to ~ 350 ps due to their now larger (22 GHz) linewidth. While not necessary in this proof-of-principle experiment, such spreading could be reduced either by using a smaller linewidth for our frequency modes (e.g., with a Fabry-Perot etalon), or by using a dispersive element with a step-wise frequency-dependent delay profile [31–33]. To verify the operation, we send in different input two-qudit states, chosen from one of 256 basis states, and measure the output after the gate. While this yields a total of $256 \times 256$ ($2^{16}$) computational input/output combinations to test, we have no active frequency-shifting elements in the SUM gate to shift 75 GHz-spaced frequencies into each other, so we make the reasonable assumption that the frequency qudit remains unchanged through the operation. This is also enforced by the high extinction ratio of the pulse shaper (~ 40 dB), which blocks unwanted frequency bins. This allows us to focus on results in the sixteen 16×16 transfer matrices measured in Fig. 4a-p (a subset with a total of $2^{12}$ input/output combinations). In each matrix, 16 different inputs with the same frequency and different time bins are sent into the SUM gate and the output time bins are measured. For this experiment, we use superconducting nanowire single photon detectors (SNSPDs), which allow us to report our data without accidental subtraction. The average computational space fidelity for the whole process, with the assumption that frequencies do not leak into each other, can be calculated as $\bar{\mathcal{F}}_C = 0.9589 \pm 0.0005$, which shows the high performance of our operation. This high fidelity benefits greatly from the high extinction ratio of the intensity modulator used to carve the time bins (~ 25 dB). To show the coherence of our SUM gate, we use this setup to perform a SUM operation on a three-dimensional input state, $|\psi\rangle_{in} = \frac{1}{\sqrt{3}}(|0\rangle_f + |1\rangle_f + |2\rangle_f)|0\rangle_t$, which results

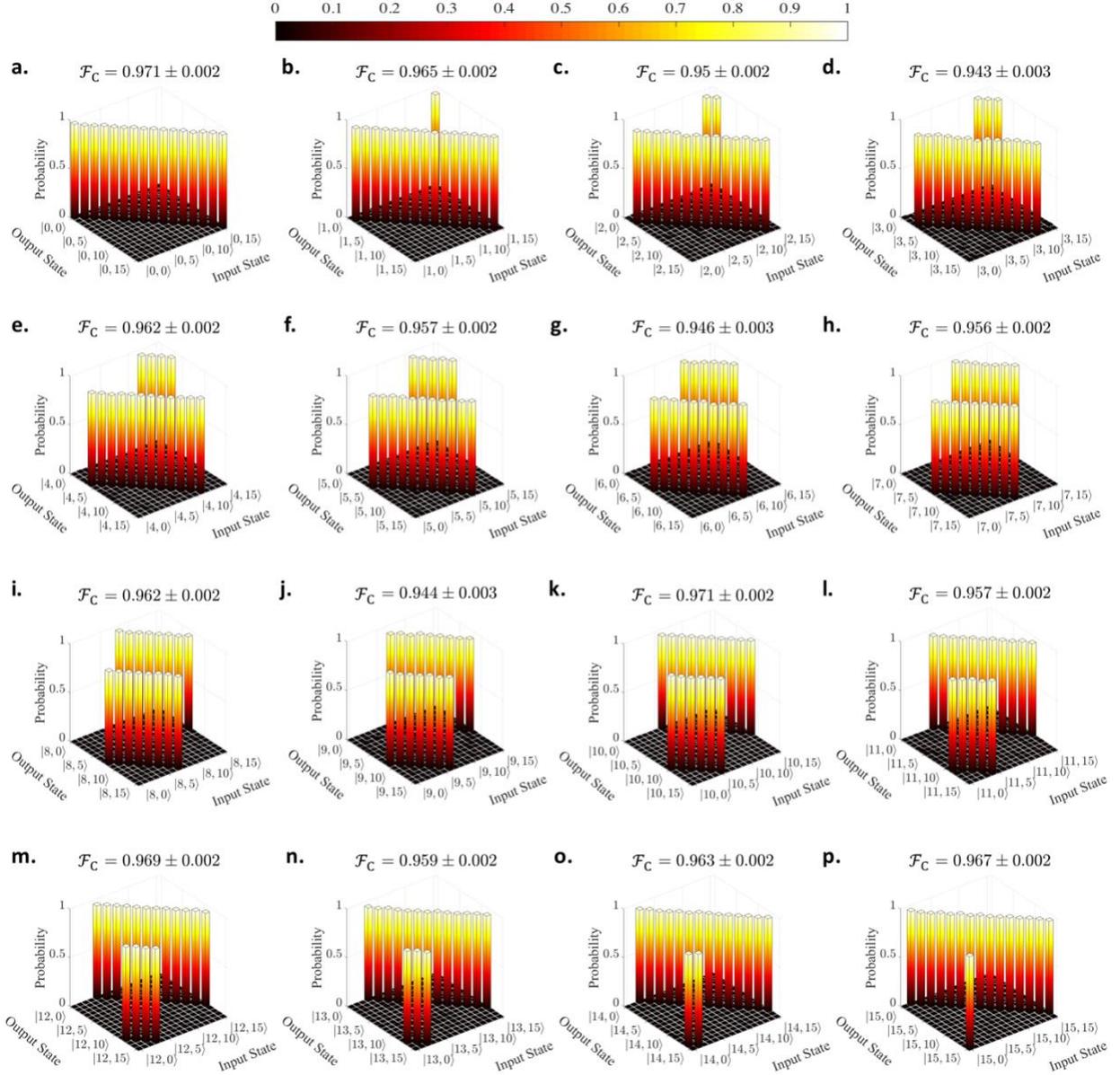

**Fig. 4. a-p,** The transfer matrices corresponding to each possible time-bin output for each individual input time bin. Each matrix is specified for one frequency input, where the matched frequency output for different time bins is measured. In $|m, n\rangle$ on the x and y axis, $m$ indicates the frequency qudit and $n$ is the time bin qudit. The computational space fidelity of each matrix is shown on top of it. Subtraction of accidentals is not employed.

in a maximally non-separable state [34] between time and frequency DoFs: $|\psi\rangle_{out} = \frac{1}{\sqrt{3}}(|00\rangle_{ft} + |11\rangle_{ft} + |22\rangle_{ft})$. To quantify the dimensionality of this state, we use an entanglement certification measure called *entanglement of formation* ($E_{of}$) [35,36]. We experimentally obtain $E_{of} \geq 1.19 \pm 0.12$ ebits, where 1 ebit corresponds to a maximally non-

separable pair of qubits, while 1.585 ebits represents the maximum for two three-dimensional parties (see Methods); in exceeding the qubit limit, our state thus possesses true high-dimensional non-separability.

One of the most crucial challenges towards optical quantum operations is the lack of on-demand photon sources. Therefore, it is interesting to consider our scheme for application to quantum communication and networking, for which operations with just a few qudits have potential impact. A gate very similar to the SUM gate is the XOR gate, which subtracts the control qudit from the target and is a requirement for qudit teleportation protocols [15]. Since teleportation of quantum states is possible using different degrees of freedom of an entangled photon pair [16], a single-photon two-qudit gate in our time-frequency paradigm could be applied directly for teleporting high-dimensional states. Specifically, the XOR gate can be demonstrated by using positive dispersion and reconfiguring the switching in the SUM gate, or in the three-dimensional case, by simply relabeling the frequency bins $|0\rangle_f \to |2\rangle_f$ and $|2\rangle_f \to |0\rangle_f$ and performing the same process as the SUM operation. Additionally, these two-qudit gates can be used for the purpose of beating the channel capacity limit for standard superdense coding for high-dimensional entangled states[37]. In such quantum communications applications for the two-qudit gates, a modest number of state manipulations brings significant value.

The demonstrated SUM gate can also be used to produce high-dimensional Greenberger-Horne-Zeilinger (GHZ) states [38]. GHZ states consist of more than two parties, entangled with each other in a way that measurement of one party in the computational basis determines the state of all the other parties [39]. It has been only recently that these states were demonstrated in more than two dimensions, where a three-dimensional three-party GHZ state was realized using the orbital angular momentum of optical states [38]. Here, we take advantage of our SUM gate and the large dimensionality of time-frequency states to generate a four-party GHZ state with 32 dimensions in each DoF. We start from the state $|\psi\rangle_{\text{in}} = \frac{1}{\sqrt{32}} |0,0\rangle_{t_s t_i} \sum_{m=0}^{31} |m,m\rangle_{f_s f_i}$, which means both signal and idler photons are initialized in the first time-bin state and are maximally entangled in the frequency domain. Then, we operate deterministic SUM gates separately on both signal and idler photons, resulting in a four-party GHZ state of the form $|\psi\rangle_{\text{out}} = \frac{1}{\sqrt{32}} \sum_{m=0}^{31} |m,m,m,m\rangle_{f_s t_s f_i t_i}$. Since the initial state only consists of the zeroth time bins, the dispersion module does not shift any of the bins outside of the computational space; hence the interferometric structure used in the full SUM gate is not required when operating within this subspace. The GHZ state is measured in the computational basis (Fig. 5); the plot contains coincidences for all basis states in the set $\{|m,n,k,l\rangle_{f_s t_s f_i t_i}; \ 0 \leq m,n,k,l \leq 31\}$. Only states whose four qudits match (i.e., $|m,m,m,m\rangle_{f_s t_s f_i t_i}$) have high counts, as expected for a GHZ state. Of course, full characterization of the state requires measurements in superposition bases as well [40], but due to the additional insertion loss associated with superposition measurements in time and frequency using interferometers and phase modulators, respectively, we were unable to measure such projections. Remarkably, the demonstrated GHZ state resides in a Hilbert space equivalent to that of 20 qubits, an impressive 1,048,576 ($32^4$) dimensions. We emphasize that the four parties of the demonstrated GHZ state are carried by only two photons, and hence cannot be used for genuine multi-partite GHZ applications such as demonstration of Bell's theorem without inequalities[39], quantum secret sharing[41], or open-destination teleportation[42]. However, the realization of such high-dimensional GHZ states indicates the potential of our time-frequency platform for quantum technologies such as near-term quantum computation and cluster-state generation[33,43].

**Discussion**

Hyper-entangled time-frequency states, as opposed to other high-dimensional optical degrees of freedom like orbital angular momentum, can be generated in integrated on-chip sources, which have gained tremendous attention in recent years due to their low cost, room temperature operation, compatibility with CMOS foundries and the ability to be integrated with other optical components. Pulse shapers, [44] phase modulators [45] and switches [46] can all be demonstrated on a chip, and a series of DWDMs and delay lines can be used to realize the equivalent functionality of on-chip CFBG. In addition, demonstration of balanced and unbalanced interferometers on-chip eliminates the need for active stabilization, which is of considerable profit for the scalability of the scheme [47]. These contributions can potentially lead to combining these sources with on-chip components designed for manipulation of these states, realizing the whole process on an integrated circuit.

High-dimensional optical states [25,26,47–49] can open the door to deterministically carry out various quantum operations in relatively large Hilbert spaces [50], as well as enable higher encoding efficiency in quantum communication protocols such as quantum key distribution [22] and quantum teleportation [16,51]. We have demonstrated deterministic single- and two-qudit gates using the time and frequency degrees of freedom of a single photon for encoding—operating on up to 256 ($2^8$)-dimensional Hilbert spaces—and carried out these gates with a high computational-space fidelity. We have shown the application of such two-qudit gates in near-term quantum computation by using them to realize a GHZ state of four parties with 32 dimensions each, corresponding to a Hilbert space of more than one million modes. Such deterministic quantum gates add significant value to the photonic platform for quantum information processing and have direct application in, e.g., simulation of quantum many-body physics[52–54].

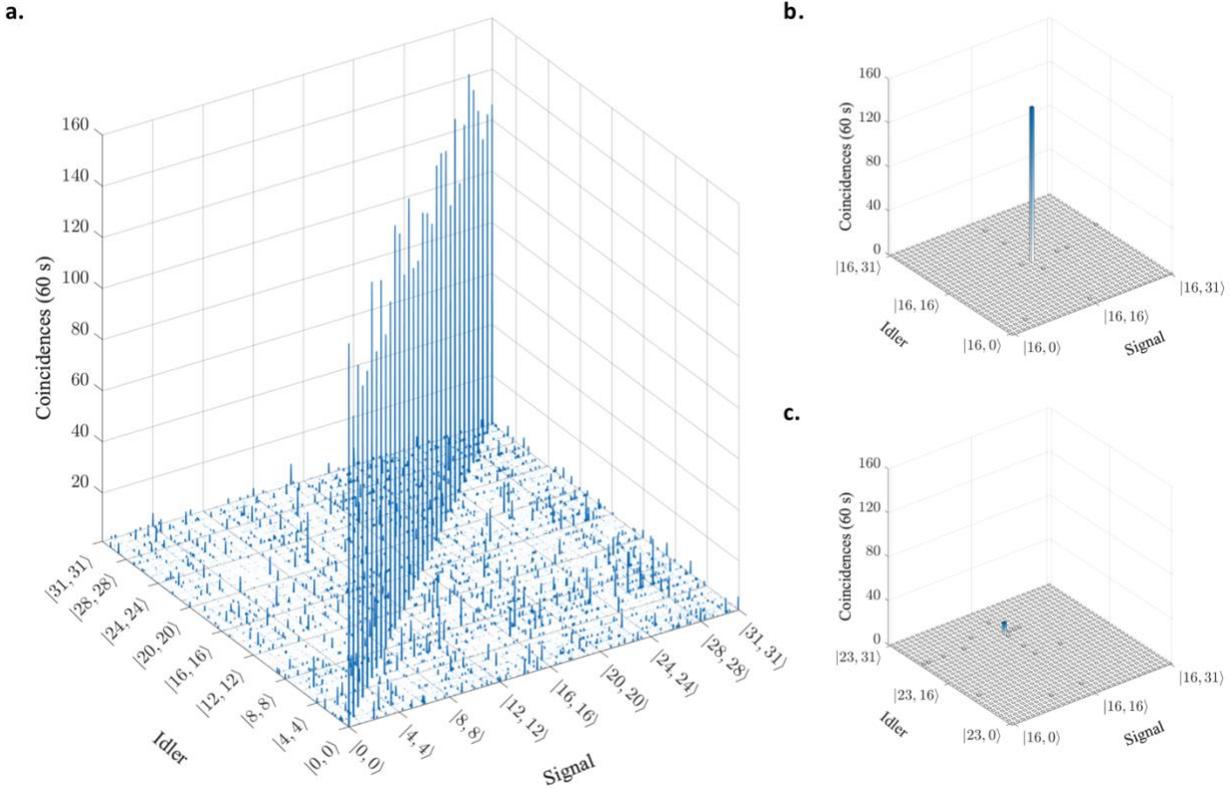

**Fig. 5. a**, Measurement of the four-party 32-dimensional GHZ state in the computational basis. The states $|m, n\rangle$ shown on the signal and idler axes correspond to frequency-bin $m$ and time bin $n$. The large coincidence peaks exist only for states with the same time-bin and frequency-bin indices for both signal and idler (32 peaks). **b-c**, Zoomed-in 32 × 32 submatrices of the matrix shown in **a**. Each submatrix shows coincidences for different signal and idler time bin indices for fixed signal and idler frequency bin indices. **b**, matched signal and idler frequency bins: large peak is observed for $|16,16,16,16\rangle_{f_s t_s f_i t_i}$. **c**, Unmatched signal and idler frequency bins. The small peak evident at $|16,16,23,23\rangle_{f_s t_s f_i t_i}$ reflects additional accidentals (multiphoton pair events) at the time bins to which frequency bins $|16\rangle_{f_s}$ and $|23\rangle_{f_i}$ are shifted. The data are shown with accidentals subtracted (coincidence to accidentals ratio of ~ 4).

**Methods**

For the time-bin single-qudit X gate shown in Fig. 2, we split the experimental setup in three stages: state preparation, X gate operation and state measurement. For the state preparation, we use an Agilent 81645A CW laser tuned to 1553.9 nm and send it into an intensity modulator (~4 dB insertion loss) and phase modulator (~3 dB insertion loss), both manufactured by EOSpace, which are used to create the time bins and control their relative phases, respectively. To implement the X gate operation, we used an MZM with two complementary outputs (~4 dB insertion loss), also manufactured by EOSpace. We use a piezo-based fiber phase shifter (General Photonics FPS-001) to control the phase difference between the two paths following the MZM. Then a 2×2 fiber coupler is used to merge the branches. For the state measurement, we used 1-bin and 2-bin delay interferometers implemented with 2×2 fiber couplers and additional piezo-based fiber phase shifters. For the time-bin X gate and computational-basis measurements of three-dimensional two-qudit gates, gated InGaAs single photon detectors (Aurea Technologies SPD_AT_M2) were used. For the rest of the measurements, we used superconducting nanowire single-photon detectors (Quantum Opus). To measure the arrival times of the photons on the single-photon detectors, a time interval analyzer (PicoQuant HydraHarp 400) is used. The stabilization of the interferometers is done by sending a CW laser at 1550.9 nm in the backwards direction and feeding the output power into a computer-based feedback loop to maintain the phase. To stabilize the X gate, we use a similar scheme with an additional circulator at the input of the gate (not shown in the figures) to retrieve the optical power in the backwards direction. The signals applied to the intensity modulators and phase modulator, as well as the trigger and synchronization signal of the single-photon detector and time interval analyzer, are generated by an electronic arbitrary waveform generator (Tektronix AWG7122B) and adjusted to the proper level by linear amplifiers.

To assess the performance of our one- and two-qudit quantum gates, we first focus on the computational-basis fidelity $\mathcal{F}_C$—one example of a so-called "classical" fidelity in the literature [55]. Defining $|n\rangle$ ($n = 0, 1, ..., N-1$) as the set of all computational basis states and $|u_n\rangle$ as the corresponding output states for a perfect operation, we have the fidelity

$$\mathcal{F}_C = \frac{1}{N} \sum_{n=0}^{N-1} p(u_n|n) \qquad (1)$$

where $p(u_n|n)$ is the probability of measuring the output state $|u_n\rangle$ given an input of $|n\rangle$. In the operations considered here, the ideal output states $|u_n\rangle$ are members of the computational basis as well, so there is no need to measure temporal or spectral superpositions in determination of $\mathcal{F}_C$. Given the measured counts, we retrieve the $N$ conditional probability distributions via Bayesian mean estimation (BME) [56,57] where our model assumes that each set of count outcomes (after accidentals subtraction) follows a multinomial distribution with to-be-determined probabilities; for simplicity, we take the prior distributions as uniform (equal weights for all outcomes). We then compute the mean and standard deviation of each value $p(u_n|n)$ and sum them to arrive at $\mathcal{F}_C$. Specifically, if $C_{u_n|n}$ signifies the counts measured for outcome $u_n$, and $C_{\text{tot}|n}$ the total counts over all outcomes (both for a given input state $|n\rangle$), BME predicts:

$$p(u_n|n) = \frac{1 + C_{u_n|n}}{N + C_{\text{tot}|n}} \pm \sqrt{\frac{1 + C_{u_n|n}}{(N + C_{\text{tot}|n})^2} \frac{N + C_{\text{tot}|n} - C_{u_n|n} - 1}{N + C_{\text{tot}|n} + 1}} \qquad (2)$$

where the standard deviation in the estimate is used for the error. Since the probabilities here each actually come from $N$ *different* distributions, we estimate the total error in $\mathcal{F}_C$ by adding these constituent errors in quadrature. Explicitly, we find $\mathcal{F}_C = 0.996 \pm 0.001$ for the X gate, $0.90 \pm 0.01$ for the CINC operation, $0.92 \pm 0.01$ for the $3 \times 3$ SUM gate, and $\bar{\mathcal{F}}_C = 0.9589 \pm 0.0005$ for the $16 \times 16$ SUM gate. The reduction in $\mathcal{F}_C$ for the two-qudit gates is due in large part to the fewer total counts in these cases, from our use of heralded single photons rather than a weak coherent state. As seen by the presence of $N$ in the denominator of Eq. (2), even when $C_{u_n|n} = C_{\text{tot}|n}$, the estimate $p(u_n|n)$ is not unity unless $C_{\text{tot}|n} \gg N$. In our experiments, the two-qudit tests have only ~100-300 total counts per input computational basis state for the $9 \times 9$ matrices (with $N$=9) and ~500-800 counts per input state for the $16 \times 16$ matrices (with $N$=16), thereby effectively bounding the maximum $p(u_n|n)$ and, by extension, fidelity $\mathcal{F}_C$. This behavior is actually a strength of BME, though, as it ensures that we have a conservative estimate of the fidelity that is justified by the total amount of data acquired [56].

While extremely useful for initial characterization, however, the computational-basis fidelity above provides no information on phase coherence. On the other hand, process tomography would offer a complete quantification of the quantum gate. Yet due to the challenging experimental complexity involved in quantum process tomography, here we choose a much simpler test which—while limited—nonetheless offers strong evidence for the coherence of our time-bin X gate. To begin with, note that all three-dimensional quantum processes can be expressed in terms of the nine Weyl operations [58]:

$$U_0 = I = \begin{pmatrix} 1 & 0 & 0 \\ 0 & 1 & 0 \\ 0 & 0 & 1 \end{pmatrix}, \qquad U_1 = X = \begin{pmatrix} 0 & 0 & 1 \\ 1 & 0 & 0 \\ 0 & 1 & 0 \end{pmatrix},$$

$$U_2 = X^2 = \begin{pmatrix} 0 & 1 & 0 \\ 0 & 0 & 1 \\ 1 & 0 & 0 \end{pmatrix}$$

$$U_3 = Z = \begin{pmatrix} 1 & 0 & 0 \\ 0 & e^{i\frac{2\pi}{3}} & 0 \\ 0 & 0 & e^{-i\frac{2\pi}{3}} \end{pmatrix}, \quad U_4 = ZX = \begin{pmatrix} 0 & 0 & 1 \\ e^{i\frac{2\pi}{3}} & 0 & 0 \\ 0 & e^{-i\frac{2\pi}{3}} & 0 \end{pmatrix}, \quad U_5 = ZX^2 = \begin{pmatrix} 0 & 1 & 0 \\ 0 & 0 & e^{i\frac{2\pi}{3}} \\ e^{-i\frac{2\pi}{3}} & 0 & 0 \end{pmatrix}$$

$$U_6 = Z^2 = \begin{pmatrix} 1 & 0 & 0 \\ 0 & e^{-i\frac{2\pi}{3}} & 0 \\ 0 & 0 & e^{i\frac{2\pi}{3}} \end{pmatrix}, \quad U_7 = Z^2X = \begin{pmatrix} 0 & 0 & 1 \\ e^{-i\frac{2\pi}{3}} & 0 & 0 \\ 0 & e^{i\frac{2\pi}{3}} & 0 \end{pmatrix},$$

$$U_8 = Z^2X^2 = \begin{pmatrix} 0 & 1 & 0 \\ 0 & 0 & e^{-i\frac{2\pi}{3}} \\ e^{i\frac{2\pi}{3}} & 0 & 0 \end{pmatrix} \quad (3)$$

The quantum process itself is a completely positive map $\mathcal{E}$ [59], which for a given input density matrix $\rho_{\text{in}}$ outputs the state

$$\rho_{\text{out}} = \mathcal{E}(\rho_{\text{in}}) = \sum_{m,n=0}^{8} \chi_{mn} U_m \rho_{\text{in}} U_n^\dagger \quad (4)$$

The process matrix with elements $\chi_{mn}$ uniquely describes the operation. The ideal three-bin X gate with process matrix $\chi_X$ has only one nonzero value, $[\chi_X]_{11} = 1$. To compare to this ideal, we assume the actual operation consists of a perfect X gate plus depolarizing (white) noise [18]. In this case we have a total operation modeled as

$$\rho_{\text{out}} = \lambda U_1 \rho_{\text{in}} U_1^\dagger + \frac{(1-\lambda)}{3} \mathbb{I}_3 \quad (5)$$

whose process matrix we take to be $\chi_N = \lambda \chi_X + \frac{1-\lambda}{9} \mathbb{I}_9$, which can be calculated by using $\mathbb{I}_3 = \frac{1}{3}\sum_{n=0}^{8} U_n \rho_{\text{in}} U_n^\dagger$ [18]. if we then assume a pure input superposition state $\rho_{\text{in}} = |\psi_{\text{in}}\rangle\langle\psi_{\text{in}}|$, where $|\psi_{\text{in}}\rangle \propto |0\rangle_t + e^{i\phi}|1\rangle_t + e^{2i\phi}|2\rangle_t$, and measure the projection onto the output $|\psi_{\text{out}}\rangle \propto |0\rangle_t + |1\rangle_t + |2\rangle_t$ (as in Fig. 2c), $\lambda$ can be estimated from the interference visibility $V$ as [60]:

$$\lambda = \frac{2V}{3-V} \quad (6)$$

and the process fidelity is then given by:

$$\mathcal{F}_P = Tr(\chi_X \chi_N) = [\chi_N]_{11} = \frac{1+8\lambda}{9} = \frac{1+5V}{9-3V} = 0.92 \pm 0.01 \quad (7)$$

as discussed in the main text.

To show the coherence of our SUM gate, we generate an input state in the signal photon which is in time-bin $|0\rangle_t$ and an equi-amplitude superposition in frequency $|\psi\rangle_{\text{in}} = \frac{1}{\sqrt{3}}(|0\rangle_f + |1\rangle_f + |2\rangle_f)|0\rangle_t$. After passing this state through the SUM gate, the time-bin state of the photon is shifted based on the frequency, leaving us with a maximally non-separable state $|\psi\rangle_{\text{out}} = \frac{1}{\sqrt{3}}(|00\rangle_{ft} + |11\rangle_{ft} + |22\rangle_{ft})$. We note that since we are starting with time-bin zero, the time bins will not fall out of the computational space; therefore, the interferometric structure is not needed for the SUM gate and a dispersion module alone can do the operation. This saves us the extra insertion loss of the interferometer, which is an important parameter due to the low photon pair rate on the detectors in this particular experiment. To measure the 3-dimensional non-separability in $|\psi\rangle_{\text{out}}$, we must vary the phases of different signal frequency bins and time bins with a pulse shaper and phase modulator, respectively. To observe the effect of this phase sweep with our relatively slow single-photon detectors (with ~ 100 ps jitter), an indistinguishable projection of all three time bins and frequency bins should be created. In general, the time bins can be projected on an indistinguishable state by using a cascade of interferometers, as illustrated in Fig. 2a. However, in our specific experiment, it is simpler to use a dispersion module with opposite dispersion to that of the module used in the SUM gate to perform the same projection. The frequency bins are then projected on an indistinguishable state using a phase modulator and pulse shaper to mix

the frequencies (Extended Fig. 1a)—a technique used previously in [26]. We note that our measurements on the signal photons are conditioned on heralding by idler frequency superposition states. To measure the interference between different signal frequency bins, the idler photons too have to be projected on an indistinguishable frequency bin using a phase modulator and pulse shaper (Extended Fig. 1a). This projection guarantees that detection of an idler photon does not give us any information on the frequency of the signal photon. Here the phases of the idler frequency bins are held constant; only the phases of the signal frequency and time bins are varied. This is in contrast to experiments in [26], where the phases of both signal and idler frequency bins were varied.

In our experiment, three-dimensional interference measurements were not possible since mixing all three frequencies together adds extra projection loss, which we cannot afford. Therefore, we vary the phases of different time bins and frequency bins to measure two-dimensional interference patterns between all three time bins and frequency bins (Extended Fig.1c). Using the visibilities of these interference patterns along with a *joint spectral intensity* (JSI) measurement (Extended Fig. 1b) can give us a lower bound on the amount of non-separability present in our system by measuring *entanglement of formation* [35,36]. The JSI denotes the correlations between the time bins and frequency bins of a signal photon heralded by an idler photon in its computational basis. This measurement was done using the same experimental setup used in Extended Fig. 1a without the equipment used for sweeping the phase of different signal time bins and projection measurements. For this measurement, the idler photons were detected after PS1, and the signal photons were detected right after the SUM gate.

Having the JSI measurement and the two-dimensional interference visibilities in hand, we have all the data needed to calculate the entanglement of formation with the assumption of having only white noise in our system, which can be expressed as:

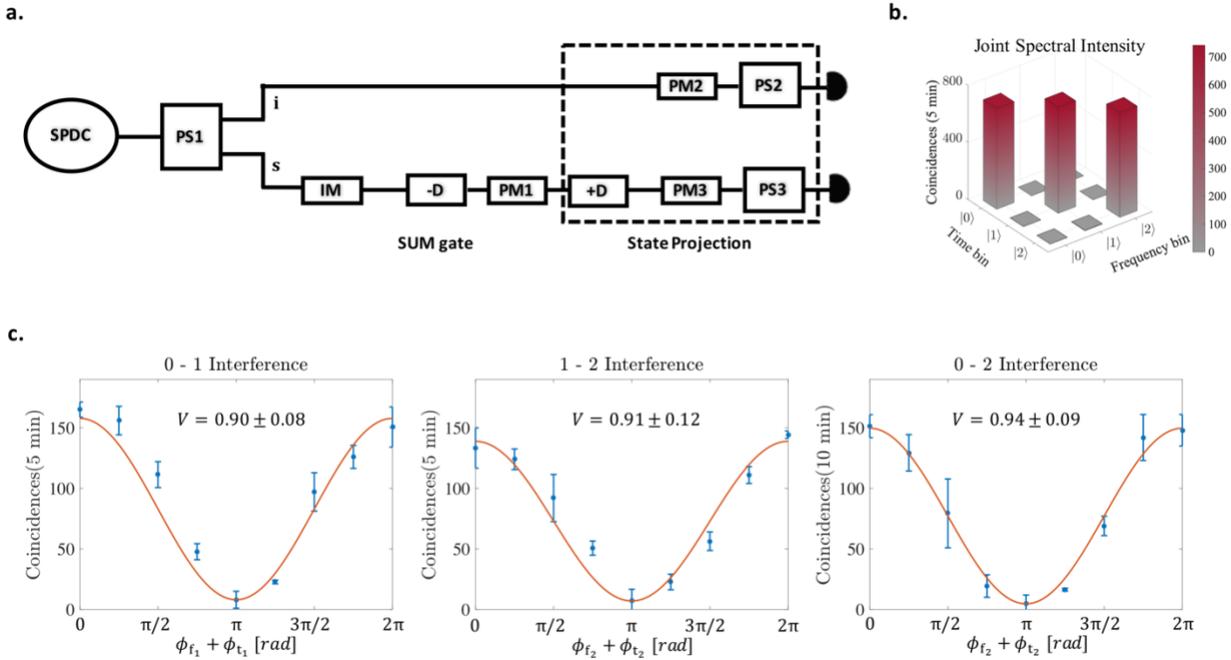

**Extended Fig. 1.** Measurement of a 3-dimensional maximally non-separable time-frequency state. **a**, The experimental setup. SPDC: spontaneous parametric down conversion, PS: pulse shaper, IM: intensity modulator, D and -D: dispersion modules with +2 ns.nm$^{-1}$ and -2 ns.nm$^{-1}$, respectively, PM: phase modulator. The same time-bin and frequency-bin spacings (1.2 ns, 75 GHz) as the 16-dimensional SUM gate experiment are used for these measurements. We note that in this experiment, the IM was placed only on the signal photons' route to avoid its insertion loss on the idler photons. **b**, Joint spectral intensity of the three-dimensional non-separable state. The accidentals were subtracted in this measurement, with a coincidence to accidentals ratio of about 30. **c**, two-dimensional interference patterns showing the coherence between all three time-frequency modes of the state. The frequency-bin and time-bin phases are varied using PS1 and PM1, respectively. Both phases are swept together from 0 to π, for a total phase sweep from 0 to 2π. The data are shown with accidentals subtracted and coincidence to accidentals ratio of about 1. Since projection of frequency bins 0 and 2 on an indistinguishable frequency bin undergoes more projection loss, the coincidences between modes 0 and 2 were measured in 10 minutes.

$$E_{\text{of}} \geq -\log_2\left(1 - \frac{B^2}{2}\right) \quad (8)$$

where

$$B = \frac{2}{\sqrt{|C|}} \left( \sum_{\substack{(j,k) \in C \\ j<k}} |\langle j,j|\rho|k,k\rangle| - \sqrt{\langle j,k|\rho|j,k\rangle\langle k,j|\rho|k,j\rangle} \right) \quad (9)$$

Here, $C$ is the number of indices $(j,k)$ used in the sum. This measurement is useful when we do not have access to all the elements of the density matrix. $\langle j,j|\rho|k,k\rangle$ ($j \neq k$) elements indicate the coherence between modes $j$ and $k$, and can be lower-bounded using the two-dimensional visibilities. The terms $\langle j,k|\rho|j,k\rangle$ can be calculated using the elements of the JSI. Using these values, we measure $E_{\text{of}} \geq 1.19 \pm 0.12$ ebits, which indicates greater than two-dimensional non-separability in our two-party system, more than one standard deviation away from the threshold.

To generate the 32-dimensional four-party GHZ state, the signal and idler go through the same dispersion module (-2 ns.nm$^{-1}$). After dispersion, the signal frequency bins farther away from the center of the spectrum are delayed more, but the idler frequency bins are delayed less as we move farther away from the center. In order to write the GHZ state in the form $|\psi\rangle_{\text{out}} = \frac{1}{\sqrt{32}} \sum_{m=0}^{31} |m,m,m,m\rangle_{f_s t_s f_i t_i}$, we label the signal time bins after dispersion 0 to 31 starting from earlier time bins (time bin 0 the earliest, time bin 31 the latest), while on the idlers, we label the time bins such that the earliest time bin is 31 and the latest time bin is 0. Another choice would be to send signal and idler through separate modules with equal but opposite dispersion, in which case we would use identical time labeling. To measure the state illustrated in Fig. 5, we individually measured coincidences for the 32 different settings of both signal and idler frequency bins ($32 \times 32$ measurements). For each of these measurements, we used our event timer to assign signal and idler time bins for each coincidence, which results in a $32 \times 32$ submatrix for each signal-idler frequency setting. Therefore, we have $32^4$ measurements in total. Two of the $32 \times 32$ time-bin submatrices are shown in Fig. 5 b-c.

We use bulk switches, dispersion modules, pulse shapers and phase modulators in out experiments, which have high insertion loss (switch: 3 dB, dispersion module: 3 dB, pulse shaper: 5 dB, phase modulator: 3 dB). Therefore, we use very bright entangled photons at the input in order to have reasonable coincidence counts on our detectors in our acquisition time. Using bright biphotons gives rise to multi-pair generation which leads to the relatively high accidental rate here.

**Data availability.**
Data and analysis codes used in this study are available from the corresponding author on request.

**Acknowledgements.**

We thank B. P. Williams for discussions on Bayesian estimation. We thank M. Hosseini, N. Lingaraju, and Nathan O'Malley for discussions. J.M.L. acknowledges support from a Wigner Fellowship at ORNL. A portion of this work was performed at Oak Ridge National Laboratory, operated by UT-Battelle for the U.S. Department of Energy under Contract No. DEAC05-00OR22725.

**Funding.** This work was funded in part by the National Science Foundation under award number 1839191-ECCS. J.M.L. acknowledges support from a Wigner Fellowship at ORNL. A portion of this work was performed at Oak Ridge National Laboratory, operated by UT-Battelle for the U.S. Department of Energy under Contract No. DEAC05-00OR22725.